\documentclass[manuscript]{aastex}

\usepackage{graphicx}
\usepackage{natbib}

\usepackage{longtable}

\shorttitle{Possibility of forming three pyrimidine bases in interstellar regions}
\shortauthors{Majumdar et al.}

\begin{document}


\title{Potential formation of three pyrimidine bases in interstellar regions}


\author{Liton Majumdar}
\affil{Univ. Bordeaux, LAB, UMR 5804, F-33270, Floirac, France
}
\affil{CNRS, LAB, UMR 5804, F-33270, Floirac, France
                    \&}
\affil{Indian Centre for Space Physics, Chalantika 43, Garia Station Road,
             Kolkata- 700084, India}

\author{Prasanta Gorai}
\affil{Indian Centre for Space Physics, Chalantika 43, Garia Station Road,
             Kolkata- 700084, India}
\author{Ankan Das}
\affil{Indian Centre for Space Physics, Chalantika 43, Garia Station Road,
             Kolkata- 700084, India}

\author{Sandip K. Chakrabarti}
\affil{ S. N. Bose National Centre for Basic Sciences, Salt Lake,
              Kolkata- 700098, India}
\affil{                    \&}
\affil{Indian Centre for Space Physics, Chalantika 43, Garia Station Road,
             Kolkata- 700084, India}
%



\begin{abstract}
Work on the chemical evolution of pre-biotic molecules remains incomplete since the 
major obstacle is the lack of adequate knowledge of rate coefficients of various 
reactions which take place in interstellar conditions. In this work, we study the
possibility of forming three pyrimidine bases, namely, cytosine, uracil and 
thymine in interstellar regions. Our study reveals that the synthesis of uracil from 
cytosine and water is quite impossible under interstellar circumstances. For the synthesis of
thymine, reaction between uracil and $\mathrm{:CH_2}$
is investigated. 
Since no other relevant pathways for the formation of uracil and thymine were available in the 
literature, we consider a large gas-grain chemical network to study the chemical evolution of cytosine 
in gas and ice phases. Our modeling result shows that cytosine would be produced in cold, 
dense interstellar conditions. However, presence of cytosine is yet to be established.
We propose that a new molecule, namely, $\mathrm{C_4N_3OH_5}$ could be observable 
in the interstellar region. $\mathrm{C_4N_3OH_5}$ is a precursor (Z isomer of cytosine) 
of cytosine and far more abundant than cytosine. We hope that observation of this precursor
molecule would enable us to estimate the abundance of cytosine in interstellar regions. 
We also carry out quantum chemical calculations to find out the vibrational as well as 
rotational transitions of this precursor molecule along with three pyrimidine bases.
\end{abstract}

\maketitle

\section{Introduction}           
The process of origin of life is yet to be known with any certainty. Till date, it is a big puzzle even to 
explain the origin of complex organic and potentially pre-biotic molecules in the
interstellar medium (ISM). As per the Cologne Database for Molecular Spectroscopy (CDMS catalog) 
\citep{mull05,mull01}, 
more than $190$ molecules have been detected in the ISM or circumstellar shells
and this number is increasing steadily. Complex molecules were detected in 
circumstellar envelopes, interstellar molecular clouds, interstellar ice, comets and 
meteorites. Examples of biologically important molecules detected in molecular clouds,
range from the amino acetonitrile ($\mathrm{NH_2CH_2CN}$) \citep{belo08} and 
proto-sugar glycolaldehyde ($\mathrm{HOCH_2CHO}$) \citep{holl00} to possible 
pyrimidine and purine base precursor like $\mathrm{HCN}$ \citep{snyd71}, 
$\mathrm{HNCO}$ \citep{joha84}, $\mathrm{NCCCH}$ \citep{ziur06}, $\mathrm{HCONH_2}$ \citep{rubi71}. 

\citet{chak00a,chak00b} computed abundances of
several pre-biotic molecules including some bases of DNA and RNA during the
collapse of proto-stellar clouds. They found that most of the molecules
are formed in the gas phase even before the star or the planets are formed. Subsequently, 
\citet{gupt11,merz14} proposed other new 
reaction schemes for the formation of adenine in interstellar region.
They computed the rate coefficients of these newly proposed reactions under interstellar circumstances.
These complex bio-molecules could be synthesized both in gas and ice phases.  
Despite several observational supports, the complete chemical composition of ISM is
yet to be determined with full confidence. In order to understand chemical composition of the ISM, 
it is required to model interstellar chemistry (gas and ice phase) by considering 
appropriate physical conditions. Recently, several authors \citep{das10,das11,das13a,das13b, 
maju12,maju13,maju14a,maju14b, sahu15,caza10, chak06a, chak06b, das08a,das08b,cupp09} 
proposed to implement grain-surface chemistry to appropriately model the chemical processes. 
Considering a gas-grain chemical model, findings of \citet{chak00a,chak00b} have been revised 
with more accurate analysis by \citet{das13a} and \citet{maju12,maju13,chak15}.
\citet{maju12,maju13} found that trace amounts of
bio-molecules (adenine, alanine, glycine etc.) could be produced during the collapsing 
phase of a proto-star. They suggested that observations of some of these precursor molecules could provide
an educated guess about the abundances of these complex molecules under interstellar conditions.

Significant indications are now present that three important interstellar 
pyrimidine bases (cytosine, uracil and thymine) existed in the pre-biotic 
Earth \citep{wang12}. Despite high potential barriers, reactions among some 
interstellar molecules may lead to the formation of cytosine, uracil and thymine. 
A large number of experiments were carried out to explain formation of complex organic 
molecules in astrophysical environments. Most of these experiments mainly focused on the 
ultraviolet photo processing of low temperature ice mixture containing $\mathrm{H_2O, 
\ CH_3OH,\ CO,\ CO_2,\ NH_3}$ etc. and subsequent warming up \citep{bern95,muno03}. 
A wide varity of organic compounds were found by analysing residues. 
\citet{nuev09,nuev12,nuev14} and  \citet{mate13} found that the addition of pyrimidine to 
$\mathrm{H_2O}$-rich binary mixtures which contain $\mathrm{NH_3, CH_3OH}$ or $\mathrm{CH_4}$ 
leads to the formation uracil, cytosine and thymine in presence of  UV irradiation. 
These photo products were formed by successive addition of $\mathrm{OH, \ NH_2}$, and $\mathrm{CH_3}$ 
groups to pyrimidine. Thymine was found to be produced only when UV dose was three 
times higher than that required to form cytosine and uracil. \citet{nuev14} obtained trace amount 
of cytosine in their experiment. Low abundance of cytosine in comparison to the uracil 
was explained due to the  low concentration of $\mathrm{NH_3}$ in the initial 
ice mixture and/or conversion of cytosine into uracil via hydrolysis. 
Hydrolysis of cytosine for the formation of uracil were also studied by 
\citet{shap99} and \citet{nels01}. They claimed that the hydrolysis of cytosine is 
responsible for the non detection of cytosine in carbonaceous chondrites.
Recently \citet{gupt13} attempted quantum chemical techniques to explore the possibility of  
cytosine formation in interstellar conditions. They proposed that some radical-radical 
and radical-molecular interactions could lead to the formation of interstellar cytosine.  
Despite overwhelmingly significant quantum chemical results, till date, any of these 
species have not been observed in the ISM. This motivates us to 
throw some lights on the status of three pyrimidine bases under interstellar circumstances.

Plan of this paper is the following. In Section 2, the computational details and 
construction of chemical model are discussed. Results are presented in Section 3, and finally, 
in Section 4, we draw our conclusions.

\section{Methods and Computational Details}

\subsection{Chemical modeling}

Our chemical model consists of a large gas-grain network where gas and grains are 
assumed to be interacting with each other continuously to exchange their chemical components. 
 
For the simplicity, here we assume a static molecular cloud, with a constant density ($n_H$) 
and constant temperature ($T$). Since, we are considering a dense cloud condition, 
we assume that gas and grains are well coupled and have the same temperature (i. e., $T_{gas}=T_{grain}=T$). 
We adopt the initial elemental abundances (Table 1) by following \citet{leun84}. 
This kind of initial elemental abundances typical of the low-metal abundances has been adopted for 
TMC-1 \citep{das15a}. 

Our present gas phase chemical network consists of $6349$ reactions between $637$ species and 
surface chemical network consists of $305$ reactions between $292$ species. Various types of 
gas phase reactions are considered in the network, namely, ion (cation)-neutral, neutral-neutral, 
charge exchange, dissociative recombination, photo-dissociation, cosmic ray induced photo-dissociation, 
radiative association, associative detachment, radiative electron attachment and mutual neutralization. 
Except molecular hydrogen (according to \citet{leit85}, sticking coefficient of $\mathrm{H_2}$ $\sim 0$ in 
low temperatures) and Helium (\citet{robe00} assumed that Helium would not stick to grains), 
depletion of all the gas phase neutral species onto the grain surface is considered with a 
sticking probability of unity.

\begin{table*}
\centering{
\scriptsize
\caption{Initial abundances used relative to total hydrogen nuclei.}
\begin{tabular}{|c|c|}
\hline
Species&Abundance\\
\hline\hline
$\mathrm{H_2}$ &    $5.00 \times 10^{-01}$\\
$\mathrm{He}$    &    $1.00 \times 10^{-01}$\\
$\mathrm{N}$     &    $2.14 \times 10^{-05}$\\
$\mathrm{O}$     &    $1.76 \times 10^{-04}$\\
$\mathrm{H_3}$$^+$&    $1.00 \times 10^{-11}$\\
$\mathrm{C^+}$ &    $7.30 \times 10^{-05}$\\
$\mathrm{S^+}$ &    $8.00 \times 10^{-08}$\\
$\mathrm{Si^+}$&    $8.00 \times 10^{-09}$\\
$\mathrm{Fe^+}$&    $3.00 \times 10^{-09}$\\
$\mathrm{Na^+}$&    $2.00 \times 10^{-09}$\\
$\mathrm{Mg^+}$&    $7.00 \times 10^{-09}$\\
$\mathrm{P^+}$ &    $3.00 \times 10^{-09}$\\
$\mathrm{Cl^+}$&    $4.00 \times 10^{-09}$\\
$\mathrm{e^-}$ &    $7.31 \times 10^{-05}$\\
$\mathrm{HD}$&  $ 1.6 \times 10^{-05}$\\
\hline
\end{tabular}}
\end{table*}

\subsection{Gas phase reaction network and rate coefficients}
For the gas-phase chemical network, we follow UMIST 2006 database \citep{wood07}. 
We considered the formation of various gas phase pre-biotic molecules following 
\citet{maju12,maju13,chak15}. In this work, we also consider various 
neutral-neutral reactions, radical-molecular reactions 
\citep{gupt13} for the formation of cytosine. 

\citet{gupt13} use quantum chemical techniques to explore the possibility of cytosine formation
by radical-radical and radical-molecule interaction schemes by barrier-less or low barrier pathways. 
They considered two different pathways (scheme 1 and scheme 2) for the formation of cytosine. 
In Table 2, we show all these pathways, where, reactions i-vii correspond to the scheme 1 
and reactions viii-xiv correspond to the scheme 2 for the production of cytosine as 
mentioned in \citet{gupt13}.
{For the formation of uracil from cytosine, we consider reaction xix, where 
reaction between water and
cytosine were considered. For the synthesis of thymine from uracil, we consider reaction numbers 
xx, where the reaction with $\mathrm{:CH_2}$ is considered. 
Here, for the simplicity and due to the unavailability of the destruction reactions, we are
not considering any destruction reactions either in the gas phase or in the ice phase and thus
predicted abundances are only the upper limit.

Since exothermic and barrier-less reactions do not follow temperature-dependent Arrhenius expression, 
an estimate of rate coefficient at $30$ K could be made on the basis of the semi-empirical relationship developed by Bates (1983): 
\begin{equation}
k = \frac{1 \times 10^{-21} A_r (6E_0 + N - 2)^{3N-7} }{(3N - 7)!}\  cm^3 \ s^{-1},
\end{equation}
where, $E_0$ is the association energy in $eV$, $A_r$ is the transition probability in $s^{-1}$ 
(assumed to be $100$), and $N$ is the number of nuclei in the complex. A limiting rate is adopted if 
the calculated value exceeds the limit set by the following equation. 
\begin{equation}
k = 7.41 \times 10^{-10} \alpha^{1/2} (10/\mu)^{1/2} \ cm^3 s^{-1},
\end{equation}
where,  $\alpha$ is the polarizability in $\AA^3$ and $\mu$ is the reduced mass of the 
reactants on the $^{12}{\mathrm{C}}$ $amu$ scale. 

Reactions, which are considered for the formation of cytosine are shown in Table 2 along with their
gas phase rate coefficients at $10$ K. All the reactions in the reaction network of cytosine 
(reactions i-xiv) are barrier-less except reaction iii. 
These reactions involve several simple neutral molecules and radicals such as 
$\mathrm{HNCO}$ \citep{bock00,buhl73,nguy91}, 
cyanoacetelyne $\mathrm{HCCCN}$ \citep{bock00, ziur06, cous99}, 
propynylidyne $\mathrm{CCCH}$ \citep{luca95}, $\mathrm{NH_2}$ \citep{vand93}, 
$\mathrm{NH}$ \citep{wage93}, $\mathrm{OCN^-}$ \citep{huds04}. 
Since reaction iii possesses a high activation barrier ($52.30$ kJ/mol, $1$ kJ/mol$=120.274$ K), 
this reaction is not feasible in the low temperature region.

\citet{gupt13} calculated activation barriers for these reactions 
but not the rate coefficients of these reactions. We used their activation barriers 
to compute rate coefficients of  these reactions by Eqn. 1. If the calculated rate exceeds 
the limiting rate ($\sim 10^{-9}$) set by Eqn. 2, we use the limiting rate only. 
For the computation of the limiting rate, we require to know the isotropic 
polarizability of the comparatively larger and stable reactant partner. 
This polarizability values were not available from \citet{gupt13}. 
Polarizability of a species is defined as the second derivative of energy with respect to 
electric field, i.e., the first derivative of dipole moment with respect to the electric filed. 
This defines the ability for a molecule to be polarized. The total isotropic mean 
molecular polarizability of a species is the mean value of polarizabilities in all 
three orthogonal directions. The polarizability tensor (is a $3\times3$ tensor) is said 
to be isotropic if its usual matrix representation is diagonal, with the three diagonal 
elements equal. 
For the computation of the isotropic polarizability, first we find out optimized geometry 
of the desired species and then by doing frequency calculation or polar calculation 
modeling on the optimized geometry, we find the isotropic polarizability. 
For this computation, we use the density functional theory based B3LYP functional with the 
6-311++G** basis set of Gaussian 09 program.  

We found that for the reactions ii, iv, v, vi, vii, x, xii and xiv, 
rate coefficients calculated by Eqn. 1 are crossing the limiting values. 
Comparatively larger and stable molecules in these reactions are
$\mathrm{C_3H_3N, \ C_4N_3H_5O,\ C_4N_3H_6O,\ C_4N_3H_6O,\ C_4N_3H_6O,\ C_4N_2H_2O,\ C_4N_3H_4O}$ 
and $\mathrm{C_4N_3H_6O}$ for  the reactions ii, iv, v, vi, vii, x, xii and xiv respectively. 
Our calculated isotropic polarizability of these species are found to be 
$6.94, \ 10.88, \ 11.31, \ 11.31, \ 11.31, \ 9.05, \ 11.37, \ 11.31$ in the unit of $\AA^3$  
for $\mathrm{C_3H_3N, \ C_4N_3H_5O,\ C_4N_3H_6O,\ C_4N_3H_6O,\ C_4N_3H_6O,\ C_4N_2H_2O,\ C_4N_3H_4O}$ 
and $\mathrm{C_4N_3H_6O}$ respectively.

\begin {figure}
\centering{
\includegraphics[height=6cm,width=10cm]{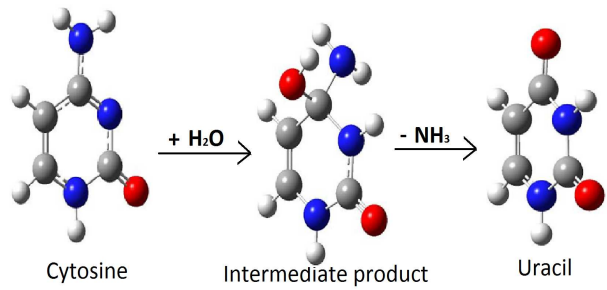}
\caption{Uracil formation in two steps: $1^{st}$ step is the addition of water with 
cytosine and $2^{nd}$ step is the elimination of ammonia from the intermediate product.}
\label{fig-1}}
\end {figure}

\citet{shap99,nels01} proposed the hydrolysis of cytosine could form uracil (reaction xix).
We investigate the usefulness of reaction xix under interstellar circumstances. 
Actually reaction xix is a two step process. First step is the water addition and $2^{nd}$ step is the
$\mathrm{NH_3}$ elimination (Fig. 1). In the $1^{st}$ step,  water reacts with cytosine to 
produce the $1^{st}$ product. We compute the change of enthalpy ($\Delta H$) for this step at 
B3LYP/6-311++G** level of theory. In gas phase, $\Delta H$ for the $1^{st}$step is $42.3123$ kJ/mol.
Thus it is highly endothermic in nature. We also calculate 
the activation barrier for the $1^{st}$ step, which is $162.5$ kJ/mol for the gas phase.
These values are quite high in relation to the interstellar circumstances 
and the process is very unlikely to go through.
In the cold interstellar clouds, averaged translational 
temperatures of the reactants are about $10$ K, which can rise up to $4000$ K in the outer 
photosphere of carbon stars \citep{kais99}. The transition state for the $1^{st}$ step (gas phase) 
is shown in Fig. 2a.

\begin {figure}
\centering{
\includegraphics[height=5cm,width=9cm]{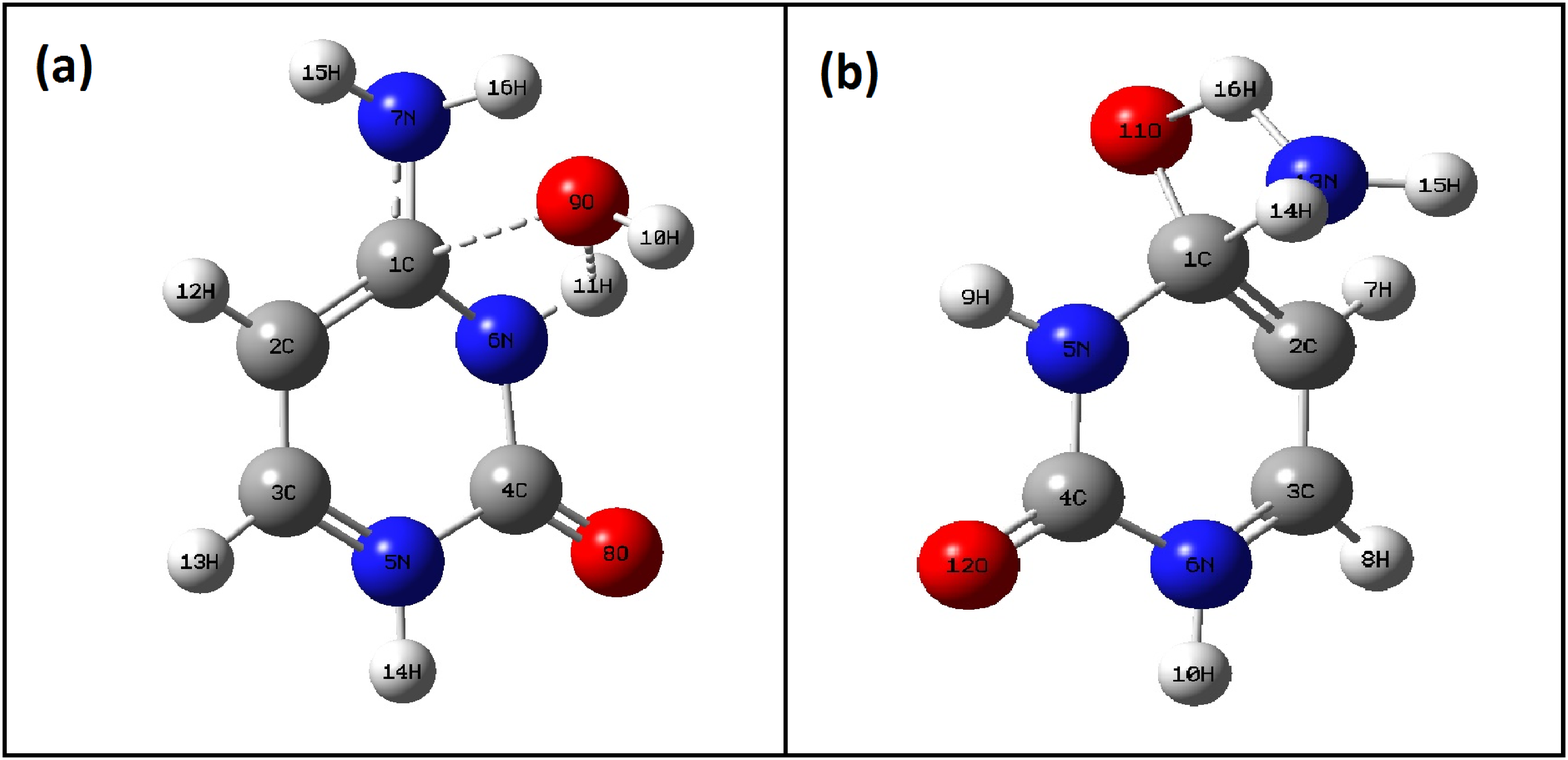}
\caption{(a) Transition state of the $1^{st}$ step, (b) Transition state of the $2^{nd}$
step of reaction xix.}
\label{fig-2}}
\end {figure}

In the $2^{nd}$ step, there is an elimination of $\mathrm{NH_3}$ from the intermediate product. 
We compute $\Delta H$ for this reaction at the same level of theory which comes out 
to be $-83.788$ kJ/mol for the gas phase. Thus, $2^{nd}$ step is highly exothermic 
in gas phase. Activation barrier for this step comes out to be $117.76$ kJ/mol for the 
gas phase. It is very interesting to note that despite highly exothermic nature of the $2^{nd}$ step, 
it possesses very high activation energy barrier in the gas phase. It is very hard to 
overcome this energy barrier under interstellar condition. The transition state for the $2^{nd}$ 
step is shown  in Fig. 2b.

It is very clear from the above discussions that under the interstellar circumstances,  
formation of uracil from cytosine is not possible by the proposed reactions. \citet{skle04} 
also investigated the conversion of cytosine to uracil under deamination reaction catalyzed 
by yeast cytosine deaminase (a zinc metalloenzyme of sufficient biomedical interest) by 
using ONIOM method. Though their studies were not related to the interstellar scenario, 
it also shows that both the rate determining steps possess  positive energy barriers. 
\citet{wang12} carried out some extensive ab-initio calculations to investigate the 
possibility  of the formation of cytosine, uracil and thymine in interstellar regions. 
Various deamination reactions of cytosine were investigated by various authors but all 
these deamination reactions are indeed very slow in nature (\citet{wang12} and references therein).  
For the formation of uracil, \citet{wang12} proposed a three body reaction between water, 
Urea and ${\mathrm:CCCO}$ which is very unlikely to happen in interstellar condition. They also 
investigated the interconversion of uracil/cytosine by using neutrals, radicals and anions 
derived from ammonia. It is required to have minimum $108$ kJ/mol energy for such interconversion. 
Thus, the chances of this interconversion are also limited.

We investigate thymine formation between uracil and $\mathrm{:CH_2}$ (reaction xx) by following 
\citet{wang12}. 
Transition state of this reaction is shown in Fig. 3. For this reaction,  
$\Delta H$ is $-456.417$ kJ/mol whereas the activation barrier is  $-84.96$ kJ/mol. 
Our studies are in excellent agreement with the previous studies by \citet{wang12}.
For the computation of the gas phase rate coefficient of reaction xx, we use
Eqn. 2. Polarizability of the largest reactant (uracil) we used is $9.85$ $\AA$.
Despite the low barrier energies of reaction xx, thymine would not be produced by this reaction. 
The reason behind is that uracil is required for reaction xx and uracil formation is limited
by the high barriers of reaction xix. Thus the proposed gas phase pathways of Table 2 would only produce 
cytosine.

\begin {figure}
\hskip 2.5cm
\includegraphics[height=9cm,width=13cm]{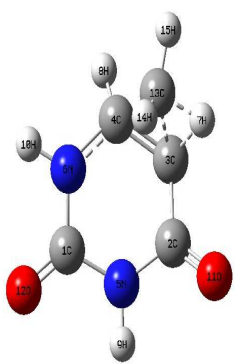}
\vskip -3.5cm
\caption{Transition state of reaction (xx).}
\label{fig-3}
\end {figure}

\begin{table*}
\scriptsize
\centering{
\addtolength{\tabcolsep}{-4pt}
\caption{Reaction network along with its rate coefficients for the formation of three pyrimidine bases at
$T=10$ K}
\vspace{1cm}
\begin{tabular}{|c|c|c|c|c|}
\hline
{\bf Species}&{\bf Reaction pathways} &{\bf Gas phase}&{\bf Rate coefficients in} &{\bf Rate coefficients in} \\
&{\bf in gas/ice phase}&{\bf activation barriers }&{\bf gas phase}& {\bf ice phase}\\
&&{\bf in (kJ/mol)} &{\bf (cm$^3$ s$^{-1}$) at $10$ $K$}&{\bf (s$^{-1}$) at $10$ K} \\
\hline
\hline
&(i) $\ \mathrm{HCCC+NH_2 \rightarrow C_3H_3N}$ & $-425.51^a$ & ${1.07 \times 10^{-9}}^c$&${2.97 \times 10^{-6}}^e$\\
&(ii)$\ \mathrm{C_3H_3N + NH \rightarrow C_3N_2H_4}$  & $-443.50^a$ &${1.37 \times 10^{-9}}^d$&${4.96 \times 10^{-3}}^e$\\
&(iii) $\ \mathrm{C_3N_2H_4+HNCO \rightarrow C_4N_3H_5O}$ & $52.30^a$ & $-$&-\\
{\bf cytosine scheme 1}&$(iv)\ \mathrm{C_4N_3H_5O +H\rightarrow C_4N_3H_6O}$  & $-326.77^a$ &${5.97 \times 10^{-9}}^d$&${4.91 \times 10^1}^e$\\
&(v)$\ \mathrm {C_4N_3H_6O +NH \rightarrow C_4H_5N_3O+NH_2}$  & $-249.37^a$ &${1.67 \times 10^{-9}}^d$&${4.96 \times 10^{-3}}^e$\\
&(vi)$\ \mathrm{C_4N_3H_6O +OH \rightarrow C_4H_5N_3O+H_2O}$  & $-361.08^a$ &${1.59 \times 10^{-9}}^d$&${1.91\times 10^{-11}}^e$\\
&(vii)$\ \mathrm{C_4N_3H_6O +NH_2 \rightarrow C_4H_5N_3O+NH_3}$  & $-314.64^a$ &${1.63 \times 10^{-9}}^d$&${2.98 \times 10^{-6}}^e$\\
\hline
&(viii)$ \ \mathrm{HCCCN + OCN^- \rightarrow C_4N_2OH}$   & $-83.68^a$ &${1.94\times 10^{-16}}^c$&${7.92 \times 10^{-14}}^e$\\
&(ix)$ \ \mathrm{C_4N_2OH + H \rightarrow C_4N_2H_2O  }$   &$ -100.42^a$ &${1.33 \times 10^{-15}}^c$&${4.91 \times 10^1}^e$\\
&(x) $\ \mathrm{C_4N_2H_2O + NH \rightarrow C_4N_3H_3O   }$   &$-449.09^a$ &${1.51 \times 10^{-9}}^d$&${4.96 \times 10^{-3}}^e$\\
&(xi) $\ \mathrm{C_4N_3H_3O + H \rightarrow C_4N_3H_4O }$   & $-248.11^a$ &${6.42 \times 10^{-10}}^c$&${4.91 \times 10^1}^e$\\
{\bf cytosine scheme 2}&(xii)$\  \mathrm{C_4N_3H_4O + H \rightarrow C_4N_3OH_5  }$   & $-304.60^a$ &${6.11 \times 10^{-9}}^d$&${4.91 \times 10^1}^e$\\
&(xiii)$\  \mathrm{C_4N_3OH_5 + H \rightarrow C_4N_3H_6O  }$   & $-157.32^a$ &${6.6 \times 10^{-13}}^c$&${4.91 \times 10^1}^e$\\
&(xiv) $\ \mathrm{C_4N_3H_6O + H \rightarrow C_4H_5N_3O + H_2  }$   & $-374.89^a$ &${6.09 \times 10^{-9}}^d$&${4.91 \times 10^1}^e$\\
&(xv) $\ \mathrm{HNCO + H \rightarrow HNCHO  }$   & -&-&${4.91 \times 10^1}^e$\\
&(xvi) $\ \mathrm{HNCO+H \rightarrow NH2CO}$   & -&-&${4.91 \times 10^1}^e$\\
&(xvii)$ \ \mathrm{HNCO+NH_3 \rightarrow OCN^-+{NH_4}^+  }$   & -&-&${8.82 \times 10^{-10}}^e$\\
&(xviii) $ \ \mathrm{HNCO + H_2O \rightarrow OCN^- + H_3O^+}$ &-&-&${9.04 \times 10^{-14}}^e$\\
\hline
{\bf Uracil}&$(xix) \ \mathrm{C_4H_5N_3O+H_2O \rightarrow Reaction \ intermediate}$ &$ 162.5^b$ &$- $&$-$\\
&$ \mathrm{Reaction \ intermediate \rightarrow C_4H_4N_2O_2+NH_3}$ &$ 117.76^b$  &$- $&$-$\\
\hline
{\bf Thymine}&$(xx) \ \mathrm{C_4H_4N_2O_2+CH_2 \rightarrow C_5H_6N_2O_2}$&$-84.96^b$&${1.61 \times 10^{-9}}^d$&${1.67 \times 10^{-7}}^e$\\
\hline
\end{tabular}}
$^a$ Gupta et al. (2013)\\
$^b$ Present work\\
$^c$ Calculated by Eqn. 1\\
$^d$ Calculated by Eqn. 2\\
$^e$ Calculated by Eqn. 3\\
\end{table*}

\subsection{Surface reaction network and rate coefficients}

It is now well established that grain chemistry plays a major role for the formation of  
complex molecules in interstellar region. Interstellar grain acts as a catalyst for the 
formation of complex interstellar molecules. In comparison with the dilute and tenuous 
interstellar gas, reaction probabilities on interstellar grains are enhanced by some orders
of magnitudes. A detailed discussions on our gas-grain chemical model are already presented in 
\citet{das10,das11,das13a,das13b,das15a,das15b,das15c,maju12,maju13,maju14a,maju14b,siva15}.

Binding energies of the surface species mainly 
dictate chemical enrichment of interstellar grain mantle. Required desorption energies 
($E_d$) for this calculations are mainly taken from \citet{alle77}, 
\citet{hase93}. Following 
\citet{hase92}, here also, we assume that the binding energy against diffusion 
($E_b$) $= 0.3E_d$ except the case of atomic hydrogen. Similar to \citet{hase92}, 
here also we use, $E_b = 100$ K for atomic hydrogen. Desorption energy of $\mathrm{HNCO}$ 
is considered to be $1425$ K  \citep{garr08}. Since desorption energy of $\mathrm{HNHCO}$, 
$\mathrm{NH_2CO}$, $\mathrm{OCN^-}$, $\mathrm{NH_4}^+$, $\mathrm{H_3O}^+$ were 
unavailable, we choose their desorption energies to be $1425$ K,
same as HNCO, so that all their binding energies are $E_b=0.3E_d \sim 427.5$ K.
For the larger species, such as $\mathrm{C_4N_2HO}$, $\mathrm{C_4N_2H_2O}$, 
$\mathrm{C_4N_3H_3O}$, $\mathrm{C_4N_3H_4O}$, $\mathrm{C_4N_3OH_5}$ \& $\mathrm{C_4N_3H_6O}$, 
we set the desorption energy to $5000$ K. Rate coefficient of the ice phase reactions are 
calculated by the following relation \citep{hase92},
\begin{equation}
 R = k_{ab} (R_a + R_b )N_a N_b n_d, 
\end{equation}
where, $N_a$ and $N_b$ are the number of $a$ and $b$  species on an average grain respectively, 
$k_{ab}$ is the probability for the reaction to happen upon an encounter and $n_d$ is dust-grain 
number density, $R_a$ and $R_b$ are the diffusion rates, for species $a$ and $b$ respectively.
Diffusion rate ($R_a = N_s t_a$ and $R_b=N_s t_b$) depends on the time required to traverse the
entire grain by the reactive species, which in turn depends on
$$
ta = 1/\nu_0 exp(E_b /kT_g ) sec,
$$
where $\nu_0$ is the characteristic vibration frequency of the adsorbed species, $E_b$ 
is the binding energy, $T_g$ is the grain 
temperature and $N_s$ is the number of surface sites on a grain. For the computation of the 
characteristic vibration frequency ($\nu_0$), we utilize the following harmonic oscillator relation:
$$
\nu_0 = \sqrt{2n_s E_d /\pi^2 m},
$$
where, $n_s$ is the surface density and $m$ is the mass of the species. As in our previous model,
here also, we use $N_s= 10^6$ and $n_s = 2 \times 10^{14}$ cm$^{-2}$ \citep{das15a}.

Various types of evaporation mechanisms are considered here, namely, thermal desorption,
cosmic ray induced desorption and non-thermal desorption. Thermal desorption time scale depends 
on the desorption energy of the species by following the relation \citep{hase92}:
$$
t_{thermal}={\nu_0}^{-1}exp(E_d/KT_g) \ sec^{-1}.
$$
Cosmic ray induced desorption time scale follows from the relation described by \citet{hase93},
$$
t_{crd}=t_{thermal}/f(70,K),
$$
where, $f(70,K)=3.16 \times 10^{-19}$.
\citet{garr07} estimated non-thermal desorption rate via
exothermic surface reactions by considering Rice-Ramsperger-Kessel (RRK) theory.
They parameterized non-thermal desorption by assuming certain approximations. They
assumed that a fraction ‘f’ of the product species in qualifying reactions could desorb
immediately and the rest (1-f) fraction remains as a surface bound product. Here, we
apply this mechanism to all surface reactions which result in a single product. Fraction
‘f’ is calculated by;
$$
f=\frac{aP}{1+aP},
$$
where, $a$ is the ratio between surface molecule bond frequency to frequency at which energy 
is lost to grain surface. Here, as in \citet{garr07}, we consider a moderate value
of $a$ (i.e., $a=0.012$) for our simulation.

\citet{gupt13} carried out quantum chemical calculations (using PCM model) for the 
formation of cytosine in ice phase as well. Their calculations shows that similar to the
gas phase reactions, only reaction iii of scheme 1 is having positive activation barrier. 
Rest of the reactions are barrier-less in ice also. Here, in our model, we are considering 
these reactions in ice phase also. Rate coefficients for these reactions in ice phase 
are computed by using eqn. 3. All the ice phase rate coefficients are shown in Table 2 at $10$ K.

For the production of cytosine via scheme 2, $\mathrm{OCN^-}$ is essential. 
The $4.62$ $\mu$m/ $2165$ cm$^{-1}$ feature is very often attributed
to the solid state $OCN^-$ \citep{soif79,misp12,grim87}. 
\citet{misp12} carried out an investigation to study the formation of $OCN^-$ 
by reaction xvii (reaction between $\mathrm{HNCO}$ \& $\mathrm{NH_3}$). According to their study,
this reaction would follow Arrhenius law with an activation energy 
of $0.4 \pm 0.1$ kJ/mol ($48 \pm 12$ K) with a pre-exponential factor of $0.0035 \pm 0.0015$ $s^{-1}$.
The proposed reaction would favourably produce $\mathrm{OCN^-}$ if it occurs in 
the $\mathrm{NH_3}$ rich (excess with respect to $\mathrm{HNCO}$) 
environment. In the ice phase, $\mathrm{NH_3}$ would be mainly 
produced by the hydrogenation of nitrogen. For the formation of $\mathrm{HNCO}$, 
we are not considering any ice phase reactions. In our gas phase network, 
$\mathrm{HNCO}$ would be produced by the reactions mentioned in \citet{mcel13}. Since, we are considering
gas-grain interactions, our ice phase would be populated by gas phase $\mathrm{HNCO}$.
\citet{misp12} showed that reaction xvii occurs in two steps. 
First step is a slow process having typical rate coefficient of 
$k_o=4.9 \times 10^{-6} \ s^{-1}$ at $10$ K and second step is relatively
faster having rate coefficient of $k_r=1.8 \times 10^{-4} \ s^{-1}$ at $10$ K. Thus overall rate
of this reaction at $10$ K is $k=k_o k_r=8.82 \times 10^{-10} \ s^{-1}$. 
In our simulation, we are also considering
this rate coefficient for reaction xvii at $10$K. 
If the rate coefficient of reaction xvii is calculated
by eqn. 3, then it would become $1.61 \times 10^{-9} \ s^{-1}$. Thus, proposed
rate coefficient by \citet{misp12} is $1.8$ times slower than the rate coefficient calculated by
thermal hopping mechanism (eqn. 3).  Reaction number xviii may also contribute to the 
formation of $OCN^{-}$ \citep{broe04}. The rate coefficient of this
reaction is calculated by eqn. 3 and the value of this rate coefficient is $9.04 \times 10^{-14} \ s^{-1}$.
It is roughly $4$ orders lower than the rate coefficient of reaction xvii. Thus $\mathrm{OCN^-}$ is
mainly produced by reaction xvii. 

We carry out quantum chemical calculations to check the possibility of 
formation of uracil (reaction xix) and formation of thymine (reaction xx) in ice phase also. 

For studying grain surface reactions i.e., reactions in astrophysical ice, one needs to 
incorporate  bulk passive influence of water.  We modelled this effect by considering a 
self-consistent reaction field in the electronic structure calculations under Gaussian 09 
program. Here the reaction components are subject to polarization effects due to the collective 
static electric field arising from the large dipole and higher moments of the bulk water matrix. 
Water changes the energetic of chemical reactions in both active and passive ways. 
Water molecules may actively serve as a catalyst, facilitating proton transfer or 
other behaviour in chemical reactions. The reaction components are also subject 
to polarization effects due to the collective static electric field arising 
from the large dipole and higher moments of the bulk water matrix. 
This is a passive influence that can also change the barrier heights of reactions 
that occur within the ice matrix. So depending on the active and passive influence 
of water, pathways, barrier as well as rates can differ. 
To study  reaction kinetics of the formation of uracil and thymine from cytosine in ice phase, 
we use density functional theory based B3LYP functional with the 6-311++G** basis set using 
Gaussian 09 program. We adopt Self Consistent Reaction Field (SCRF) modeling using the polarizable 
continuum model  (PCM) with the integral equation formalism variant (IEFPCM) as the default SCRF 
method. This method in Gaussian 09 program is used to include the bulk solvation effect of the 
interstellar grain mantle (mostly water ice). For the SCRF modeling, bulk solvent medium  is simulated 
as a continuum of the dielectric constant (for water =$78.5$) which surrounds a solute cavity, 
defined by the union of a series of interlocking spheres centered on the reactive atoms. 

As like the gas phase (Fig. 1), reaction xix would process in two steps. 
$\Delta H$ for the first step of this reaction is $76.27$ kJ/mol and for the second step is
$-95.23$ kJ/mol. Thus, first step of reaction xix is endothermic whereas the second step is 
exothermic in ice phase. Activation barriers for these two steps are $163.05$ kJ/mol  
and $108.87$ kJ/mol in the ice phase. Due to these high barrier energies,
reaction xix would not able to process in ice phase as well. Thus formation of uracil in
ice phase is not possible.
For the thymine formation (reaction number xx ) in ice phase, reaction of uracil with 
$\mathrm{:CH_2}$  is studied. $\Delta H$ for this reaction $-415.59$ kJ/mol in ice phase. Thus this 
reaction is exothermic in nature. 
Activation barriers for this reaction is 
$-65.48$ kJ/mol. Since the activation barrier for reaction
between uracil and $\mathrm{:CH_2}$ (reaction xx) is favourable under interstellar condition, 
this reaction would process in interstellar ice. Now, reaction xx require uracil for the
production of thymine. Since uracil is not forming on the ice, production of thymine on the
interstellar ice is questionable.
Thus, we are not considering reaction numbers xix-xx in our surface chemical network.
Rate coefficient for reaction xx in ice phase is shown in Table 2 for the illustration purpose.

\section{Results and Discussion}

\subsection{Modeling results}
In Fig. 4, we show chemical evolution of cytosine
in gas phase and ice phase. `Y' axis of Fig. 4 
represents logarithmic abundance of species ($n_x$) with respect to number density of 
total hydrogen nuclei in all forms ($n_H$). To mimic dense interstellar conditions, 
we consider an intermediate dense cloud ($n_H=10^5$ cm$^{-3}$) having $T=10$ K and $A_V=10$. 
$\mathrm{OCN^-}$ is the basic requirement for the formation of cytosine and it is producing
efficiently on the ice phase.  Due to the moderate desorption
energy of $\mathrm{OCN^-}$, it is populating the gas phase to process cytosine 
formation in gas phase as well.

\begin {figure}
\centering{
\includegraphics[height=8cm,width=8cm]{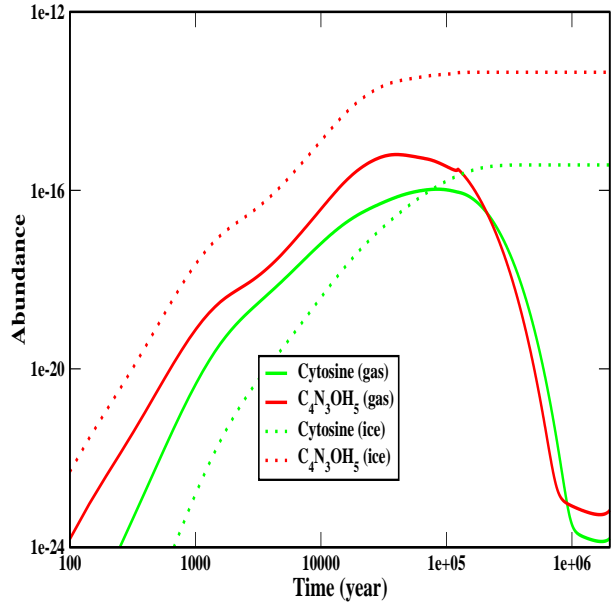}
}
\caption{Chemical evolution of cytosine along with its precursor ($\mathrm{C_4N_3OH_5}$).}
\label{fig-4}
\end {figure}

It is evident from Fig. 4 that the abundance of cytosine is beyond the present observational limit
\citep{agun13}. 
Thus it would be more helpful, if we identify an intermediate molecule as the precursor
of cytosine. Here, we consider $\mathrm{C_4N_3OH_5}$ (Z isomer of cytosine) as the precursor (Fig. 5a). 
\citet{gupt13} obtained this structure during the synthesis of cytosine starting from 
$\mathrm{OCN^-}$ and $\mathrm{HCCCN}$. There is  another isomer of cytosine (i.e., E isomer) which would also
be formed starting from $\mathrm{OCN^-}$
and $\mathrm{HCCCN}$. Structure of the E isomer of cytosine is shown in Fig. 5b. Though the 
E isomer is more stable than the Z isomer (optimization energy of the Z isomer is $6.8$ kJ/mol 
higher than the E isomer), we are considering the Z isomer only
because the pathways with this isomer are available for the formation of cytosine. We also check the most 
economic route for the formation of cytosine starting from the precursor (Z or E isomer of cytosine). 
We compare $\Delta H$ of reaction (xiii), 
where, $\mathrm{C_4N_3H_6O}$ is forming by the reaction between the precursor (Z or E isomer) 
and $\mathrm{H}$ atom. $\Delta H$ for the reaction between H atom and Z isomer is
$-155.914$ kJ/mol and between H atom and E isomer is $-154.07$ kJ/mol. Thus reaction with the 
Z isomer is more endothermic than the reaction with the E isomer towards the final step of the cytosine
production.
In Fig. 4, time evolution of $\mathrm{C_4N_3OH_5}$ is also shown. 
It is clear that it 
could efficiently be produced. 
Peak abundance of gas phase cytosine and $\mathrm{C_4N_3OH_5}$ is found to be
$1.06 \times 10^{-16}$, $6.36 \times 10^{-16}$ respectively and for the ice phase it is 
found to be $3.72 \times 10^{-16}$, $4.44 \times 10^{-14}$
respectively. It is expected that all the ice phase species would populate the gas phase during
the warm up phase of a collapsing cloud. Thus the ice phase abundance of the precursor molecule 
would be reflected in the gas phase as well.

\begin {figure}
\vskip 3cm
\centering{
\includegraphics[height=5cm,width=8cm]{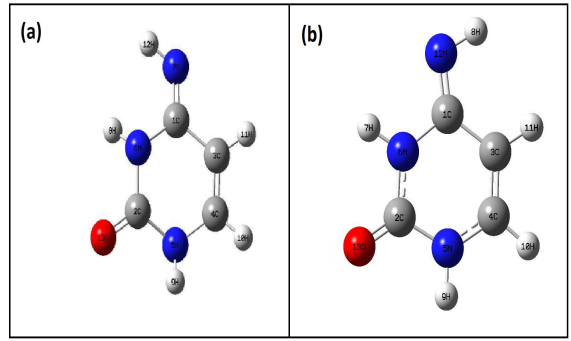}}
\caption{Structure of (a) Z isomer and (b) E isomer of $\mathrm{C_4N_3OH_5}$.}
\label{fig-5}
\end {figure}

\subsection{Astronomical Spectroscopy}

\subsubsection{Vibrational transitions}

Gas phase vibrational frequencies of three pyrimidine bases and one of their precursor 
($\mathrm{C_4N_3OH_5}$) are shown in Fig. 6. For these calculations, we use 
B3LYP/6-311G++(d,p) level of theory in Gaussian 09 program. Three most distinct peaks of cytosine 
appear at $1768$, $1680$ and $1632$ cm$^{-1}$ respectively. In case of uracil, two
peaks are very strong, which are at $1800$ and $1768$ cm$^{-1}$. In case of thymine,
two strongest peaks appear at $1800$ and $1752$ cm$^{-1}$.
In case of the precursor molecule ($\mathrm{C_4N_3OH_5}$), two strong peaks appear 
at $2359$ and $1792$ cm$^{-1}$.  For a better illustration, we show the
peak positions along with the band assignments of these species in Table A1. 
In Table A1, we compared our calculated vibrational frequencies with the 
earlier existing experimental works to justify the accuracy of our calculations. 
It is true that the validity of the band assignments are not obvious 
because there are some earlier works which reported the similar issues. But 
to the best of our knowledge, there are no such information available for $\mathrm{C_4N_3OH_5}$. 
For the sake of completeness and to validate the accuracy of our results on $\mathrm{C_4N_3OH_5}$, 
we summarize the vibrational frequencies and band assignments of 
cytosine, uracil and thymine along with $\mathrm{C_4N_3OH_5}$.

\begin {figure}
\vskip 2cm
\centering{
\includegraphics[height=8cm,width=9cm]{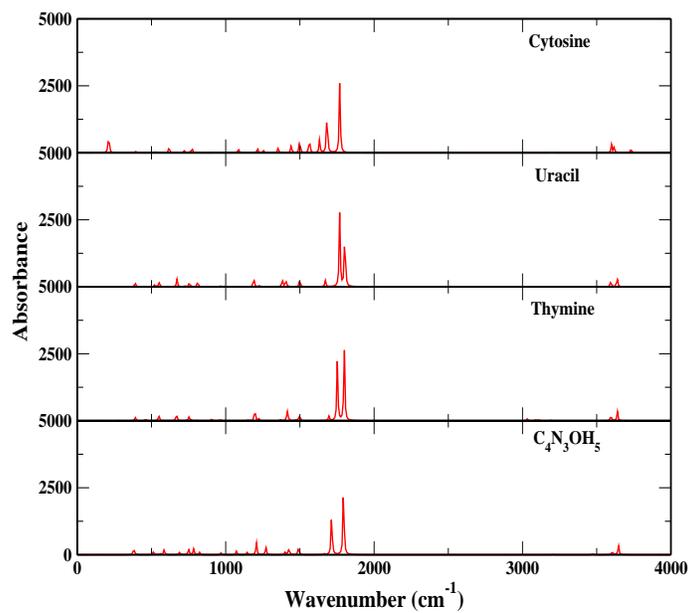}}
\caption{Infrared spectra of three pyrimidine bases and their precursor in gas phase.} 
\label{fig-6}
\end {figure}

\subsubsection{Rotational transitions}
In Table 3, rotational constants of three pyrimidine bases along with one
precursor are shown. 
It is now well known that density functional theory can be applied to various astrophysical 
problems with sufficient accuracy
\citep{rung84,pule10,piev14}. 
Accuracy of the computed spectroscopic constant depends on the choices of the models. For example,
\citet{carl13,carl14} fitted the experimental rotational 
frequencies to obtain spectroscopic constants and compared it with the 
theoretical quantum chemical calculations at the B3LYP/cc-pVTZ level of theory 
using Gaussian 09 program. Their results are in excellent agreement with the 
theory. Here also, we used the B3LYP/cc-pVTZ level of theory for the computation. 
It is clear from Table 3 that our calculated rotational constants ($B$ and $C$) are in 
very good agreement with the experimentally measured values in experiments.  But experimentally measured 
value of $A$ deviates more from our calculated value. This difference could be 
attributed due to the difference between the techniques involved. Calculation of 
rotational constants refer to an equilibrium geometry, while the measured ones are 
subject to some vibrational effects also \citep{csas89}. 
Experimentally extracted values might contain some uncertainties. 
Uncertainty is basically the expected errors of the experimental frequencies 
which are based on the propagation of errors estimated from a least squares fit 
of the observed frequencies to a model Hamiltonian. These predicted uncertainties of 
transition frequencies are strongly model dependent. Predicted uncertainties are 
smaller if large number of transition frequencies are measured and then implemented in the fit.
Gaussian 09 program generate 
these parameters from the computation of anharmonic frequencies and analytic 
second derivative of energies at displaced geometries. These analytic computations 
are highly accurate and thus they does not reflect any uncertainties in their results. 
{Some earlier theoretical/experimental works have been performed to find out the rotational 
constants of cytosine and uracil. Here, we also use the B3LYP/cc-pVTZ level of theory in 
Gaussian 09 program for the computation. In Table 3, we compare our results with the existing results. 
It is clear from the table that our calculated values are in good agreement with the earlier results.
}

These rotational constants 
are used in SPCAT program \citep{pick91} to generate several rotational
transitions. These transitions are used in ASCP program \citep{kisi98,kisi00} 
for the representation of the rotational stick diagram (Fig. 7) at $300$ K.

\begin{table*}
\centering{
\scriptsize
\addtolength{\tabcolsep}{-4pt}
\caption{Rotational constants for cytosine, uracil, thymine and one of
its precursor at B3LYP/cc-pVTZ level of theory}
\begin{tabular}{|c|c|c|c|}
\hline
{\bf Species}&{\bf Parameter}&{\bf Values (in MHz)}&{\bf Experimental/observational results}\\
\hline
&$A$&$3846.1573$&$3871.54618^d$\\
{\bf Cytosine}&$B$&$2012.0252$&$2024.97804^d$\\
&$C$&$1321.6224$&$1330.33627^d$\\
\hline
&$A$&$3902.4459$&$3883.873021^a$,$3883.87825^b$,$3902.4^c$\\
{\bf Uracil}&$B$&$2022.1182$&$2023.732581^a$, $2023.73267^b$, $2022.2^c$\\
&$C$&$1331.9473$&$1330.928108^a$, $1330.92380^b$, $1332.0$\\
\hline
&$A$&$3212.0066$&$ 3201.20^e$\\
{\bf Thymine}&$B$&$1404.9181$&$ 1401.81^e$\\
&$C$&$9832.6168$&$9826.1^e$\\
\hline
&$A$&$3881.84$&\\
{\bf $\mathrm{C_4N_3OH_5}$}&$B$&$2011.83$&-\\
&$C$&$1325.08$&\\
\hline
\multicolumn{4}{|c|}{$^a$ Experimental values by \citet{brun06}}\\
\multicolumn{4}{|c|}{$^b$ Experimental values by \citet{brow88}}\\
\multicolumn{4}{|c|}{$^c$ Calculated values by \citet{brun06} at B3LYP/cc-pVTZ level of theory}\\
\multicolumn{4}{|c|}{$^d$ Calculated values by \citet{alon13} at MP2/6-311++G(d,p) level of }\\
\multicolumn{4}{|c|}{$^e$ Experimental values by \citet{brow89}}\\
\hline
\end{tabular}}
\end{table*}

\begin {figure}
\vskip 3cm
\centering{
\includegraphics[height=8cm,width=9cm]{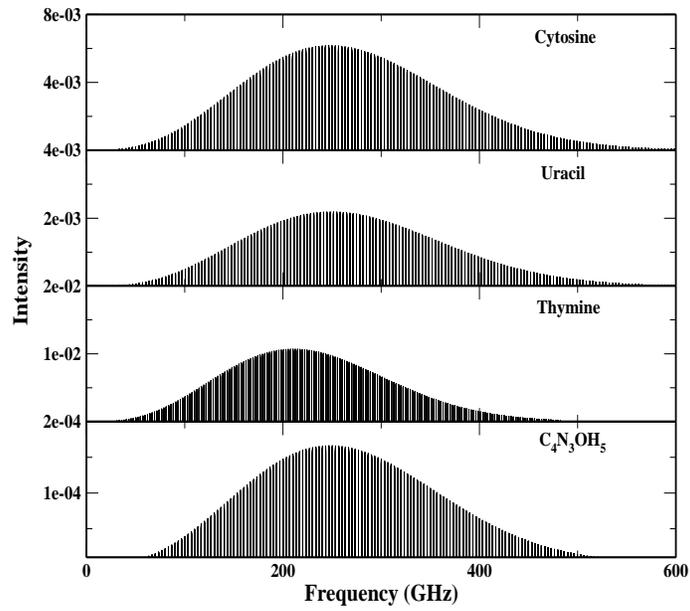}}
\caption{Rotational stick diagram of three pyrimidine bases and their precursor.} 
\label{fig-7}
\end{figure}

\section {Concluding Remarks}

In this paper, we explored the possibility of formation of three pyrimidine bases, namely, 
cytosine, uracil and thymine in interstellar region. 
We studied the hydrolysis of cytosine for the formation of uracil. 
Our results strongly suspects the validity of this reaction around the low 
temperature regime. Since uracil production is limited by the hydrolysis reaction, 
production of thymine by the reaction between uracil and $\mathrm{CH_2}$ is highly unlikely.
However, the formation of cytosine is energetically favourable (in gas and ice phase) and thus we consider
cytosine formation reactions in our gas-grain chemical model. 
Our calculated abundance suggests that trace amount of cytosine could be produced in the 
interstellar circumstances. Since, we are not considering any destruction reactions of cytosine, 
our predicted value merely reflects its  upper limit. Despite this, our calculated 
abundance of cytosine is below the observational limit. This prompted us 
to propose to observe one precursor molecule of cytosine (intermediate product of cytosine route) 
whose abundance is much greater than cytosine. $\mathrm{C_4N_3OH_5}$ could be used as 
the precursor molecule of cytosine. 
It is not obvious that the cytosine is harder to detect as compare to $\mathrm{C_4N_3OH_5}$ 
based on the abundance argument only. \citet{das15a,das15b} already discussed this issue in 
detail and also mentioned the necessity of chemical modeling coupled with spectroscopic 
calculations prior to any astronomical survey. Our computed rotational spectroscopic data, 
abundances could used for radiative transfer modeling. It is now well known how the results 
of radiative transfer modeling play a crucial role for observing any species in the ISM 
\citep{cout14}. 
Moreover, we carried out quantum chemical calculations to represent vibrational transitions 
of three pyrimidine bases along with the precursor of cytosine.
Band assignments and a comparison with the earlier works have been carried out. Rotational constants of 
these bio-molecules are also computed by quantum chemical calculations and rotational
stick diagrams are presented for the illustration purposes.

\section{Acknowledgment}
AD, SKC, \& PG thank ISRO respond project (Grant No. ISRO / RES / 2/ 372/ 11-12) and 
DST project (Grant No. SB/S2/HEP-021/2013) for partial supports. LM thanks MOES and ERC 
starting grant (3DICE, grant agreement 336474) for funding during this work.
We thank Mr. R. Saha for 
his help in data reduction. 

{}

\clearpage
\centering{\scriptsize \bf TABLE A1}\\ 
\noindent{\scriptsize VIBRATIONAL FREQUENCIES OF CYTOSINE, URACIL, THYMINE and $\mathrm{C_4N_3OH_5}$ 
IN GAS PHASE AT B3LYP 6-311g++(D,P) LEVEL OF THEORY}\\
\begin{longtable}{|c|c|c|c|c}
\hline
\centering
\addtolength{\tabcolsep}{14pt}
{\bf Species}&{\bf Peak }&{\bf Integral}&{\bf Band}&{\bf Experimental/}\\
{}&{\bf positions }&{ \bf absorption}&{\bf assignments}&{\bf calculated}\\
{}&{\bf (Gas phase)}&{\bf coefficient in}&{}&{\bf values}\\
&{\bf (in cm$^{-1}$)}& {\bf (cm  molecule$^{-1}$)}& &{\bf (in cm$^{-1})$}\\
\hline
& 128.49 & 3.33$\times$10$^{-19}$ & skeletal deformation& $270^e$, $235^f$\\
& 196.35 & 1.27$\times$10$^{-18}$ & $NH_2$ bending out of plane & $291^f$\\ 
& 211.91 & 3.85$\times$10$^{-18}$ & $NH_2$ bending&$305^f$\\ 
& 358.27 & 5.16$\times$10$^{-19}$  & $NH_2$ bending &$558.6^e$, $550^f$\\ 
& 395.08 & 3.44$\times$10$^{-18}$ & $NCC$ bending &$397^e$, $400^f$\\
& 524.64 & 2.02$\times$10$^{-18}$ & $NH_2$ twisting &$531^e$, $535^f$\\
& 532.99 &  4.16$\times$10$^{-19}$ & skeletal deformation& $531^e$, $535^f$\\
& 545.16 & 4.58$\times$10$^{-19}$ & skeletal deformation&$531^e$, $535^f$\\
& 578.70 & 4.22$\times$10$^{-19}$ & skeletal deformation&$571^e$, $575^f$\\
& 618.98 & 1.10$\times$10$^{-17}$&  NH bending &$614^e$, $614^f$\\
& 722.56 & 4.95$\times$10$^{-18}$ &C$H$ bending&$717^e$, $716^f$\\
& 762.36 & 1.54$\times$10$^{-18}$ & skeletal deformation &$749^e$, $747^f$\\
& 765.66 & 7.65$\times$10$^{-19}$& ring deformation&$767^e$, $747^f$\\
& 773.85 & 7.65$\times$10$^{-18}$ & $NCN$ bending out of plane &$784^e$, $818^f$\\
& 919.63 & 7.47$\times$10$^{-19}$  & $NC$ stretching&$2922^e$, $2930^f$\\
& 956.98 &1.09$\times$10$^{-19}$ & $CH$ bending &\\
& 986.25 &  7.60$\times$10$^{-20}$ & $CC$ stretching &\\
& 1085.37 & 7.97$\times$10$^{-18}$ &$NH_2$ rocking&  $1088^f$\\ 
{\bf Cytosine}& 1125.37 & 5.55$\times$10$^{-19}$& $CH$ bending &$1103^e$, $1090^f$ \\
& 1214.06 & 8.21$\times$10$^{-18}$ & $CH$ waging & $1198^e$, $1196^f$\\
& 1253.75 & 5.01$\times$10$^{-18}$& $NC$ stretching&$1237^e$, $1244^f$ \\
& 1353.84 &9.42$\times$10$^{-18}$  & $CH$ $NH$ rocking & $1340^e$, $1337^f$\\
&1441.96 &1.44$\times$10$^{-17}$ & $NH$ rocking& $1423^e$, $1423^f$  \\
& 1498.97 & 2.56$\times$10$^{-17}$ & $NC$ stretching &$1475^e$, $1475^f$\\
& 1564.29 & 2.79$\times$10$^{-17}$ & $CC$ stretching &$1540^e$, $1539^f$\\
& 1632.35 & 2.27$\times$10$^{-17}$ &  $NH_2$ scissoring &$1602^e$, $1598^f$\\
& 1683.00 & 8.38$\times$10$^{-17}$ & C=C stretching&$1659^e$, $1656^f$\\
& 1768.89 & 1.29$\times$10$^{-16}$ & C=O stretching&$1730^e$, $1733^f$\\
& 3192.84 & 4.31$\times$10$^{-19}$ & CH stretching&\\
& 3218.01 & 3.65$\times$10$^{-19}$  & CH stretching&\\
& 3599.58 & 1.49$\times$10$^{-17}$  & $NH_2$  symmetric stretching&$3457^e$, $3441^f$\\ 
& 3617.97 &1.17$\times$10$^{-17}$  & NH stretching& $3474^e$, $3472^f$\\
& 3731.43 & 8.41$\times$10$^{-19}$ & $NH_2$ antisym. stretching&$3575^e$, $3564^f$\\ 
\hline\hline
& 142.68 &1.45$\times$10$^{-19}$   & CN bending out of plane  &\\
& 162.29 & 4.80$\times$10$^{-20}$   & CNC bending out of plane &\\ 
& 388.06 & 3.42$\times$10$^{-18}$  & C=O bending&$374^a$, $391^b$  \\ 
& 390.53 & 4.54$\times$10$^{-18}$ & CCN bending &$395^a$, $393^b$ \\ 
& 521.16 & 3.42$\times$10$^{-18}$ & skeletal deformation &$512^a$, $516^b$, $516.5^c$\\
& 542.13 &1.37$\times$10$^{-18}$& skeletal deformation & $536.4^b$, $536^c$\\
& 551.96 & 6.92$\times$10$^{-18}$ & NH bending &$545^a$, $551.2^b$, $550^c$\\
& 559.47 & 6.91$\times$10$^{-19}$  & skeletal deformation & $559^b$, $559^c$\\
& 670.10 & 1.42$\times$10$^{-17}$  & NH bending &$660^a$, $662.2^b$, $662^c$ \\
& 724.44 & 1.61$\times$10$^{-18}$& CH bending &$717^a$, $717.4^b$, $717^c$\\
& 755.15 & 8.00$\times$10$^{-18}$ & C=O bending out of plane  &$757^a$, $756.5^b$, $757^c$ \\
& 767.54 & 5.81$\times$10$^{-19}$& ring deformation &\\
& 811.05 &9.97$\times$10$^{-18}$ & CH bending &$802^a$, $804^b$, $804^c$ \\
& 963.84 &1.857$\times$10$^{-18}$ & ring stretching &$952^a$, $958.3^b$, $958^c$ \\
& 965.78 &1.19$\times$10$^{-19}$& CH bending &$972^a$, $981.5^b$, $982^c$ \\
& 989.54 & 1.17$\times$10$^{-18}$ & ring deformation &$990^a$, $987.5^b$, $987^c$ \\
{\bf Uracil}& 1084.64 & 8.55$\times$10$^{-19}$ & NH, CH bending &$1082^a$, $1075.5^b$, $1076^c$\\
& 1188.31 & 1.76$\times$10$^{-17}$  & NC stretching  &$1185^a$, $1192^b$, $1176^c$\\
& 1226.63 &2.68$\times$10$^{-18}$   & CH bending &$1356^a$, $1217.4^b$, $1217^c$\\ 
& 1381.18 & 1.40 $\times$10$^{-17}$& CH bending & $1387^a$, $1388.7^b$, $1389^c$\\
& 1404.87 & 1.31$\times$10$^{-17}$& NH bending&$1400^a$, $1399.6^b$, $1399^c$ \\
& 1419.96 &8.85$\times$10$^{-19}$ &CH bending& \\
& 1498.28 &1.38$\times$10$^{-17}$ & NC stretching & $1515^a$, $1517^b$, $1472^c$\\
&1671.13 & 1.16$\times$10$^{-17}$& C=C stretching& $1641^a$, $1644^b$, $1644^c$\\
& 1767.68 & 1.16$\times$10$^{-16}$ & C=O stretching &$1756^a$ ,$1757.5^b$, $1758^c$\\
& 1861.92 &1.09$\times$10$^{-16}$ & C=O stretching &\\
& 3203.31 &4.44$\times$10$^{-19}$  & CH stretching &\\
& 3243.52 & 1.86$\times$10$^{-19}$ & CH stretching& $3124^a$\\
& 3596.94 &1.11$\times$10$^{-17}$ & NH stretching & $3436^a$  \\
& 3638.27 & 1.77$\times$10$^{-17}$  & NH stretching &$3484^a$\\
\hline\hline
& 107.53 & 9.79$\times$10$^{-22}$   & $CH_3$ bending &\\
& 141.81 & 1.12$\times$10$^{-19}$    & $CH_3$ torsion &\\ 
& 145.54 &  6.63$\times$10$^{-20}$   & $CNC$ bending&\\ 
& 276.56 &  4.40$\times$10$^{-19}$  & $CH_3$ bending &\\ 
& 290.10 &  2.83$\times$10$^{-20}$  & C-C=C bending &\\
& 389.43 &  3.25$\times$10$^{-18}$ & O=C bending &\\
& 391.08 & 3.55$\times$10$^{-18}$   &CCN bending &$391^c$ \\
& 460.72 &  3.17$\times$10$^{-18}$  & skeletal deformation& $462^a$, $455^c$ \\
& 545.85 &  1.27$\times$10$^{-18}$   & skeletal deformation&$541^a$, $545^d$, $540^c$\\
& 549.47 & 9.42$\times$10$^{-18}$ &  NH bending  &$541^a$, $541^d$, $545^c$\\
& 606.37 &  2.17$\times$10$^{-19}$  & skeletal deformation &\\
& 668.18 &  1.42$\times$10$^{-17}$ &  NHbending &$689^a$, $662^d$, $662^c$\\
& 732.66 &  8.90$\times$10$^{-19}$   & CC stretching &\\
& 753.41 &  7.54$\times$10$^{-18}$  &N-C=N bending out of plane &$755^a$, $754^d$, $754^c$\\
& 764.83 &  1.84$\times$10$^{-18}$   &C=O bending  &$767^a$, $764^d$, $764^c$\\
& 803.19 &  8.15$\times$10$^{-19}$  &skeletal deformation  &$804^a$, $800^c$\\
& 906.85 & 3.09$\times$10$^{-18}$   & $CH$ bending& \\
& 963.85 &  2.26$\times$10$^{-18}$    &$NC$ stretching&$963^a$, $959^d$, $959^c$\\ 
& 1022.22 & 3.59$\times$10$^{-19}$  & $CH_3$ bending & $1031^a$, $1005^d$, $1002^c$ \\
{\bf Thymine}& 1066.76 & 1.78$\times$10$^{-19}$  & $CH-3$ bending& $1078^a$, $1087^d$, $1046^c$\\
& 1148.81 & 1.23$\times$10$^{-18}$  & $NC$ stretching& \\
& 1196.20 &  2.34$\times$10$^{-17}$ & $NC$ stretching& \\
&1222.69 & 3.96$\times$10$^{-18}$  &$CC$ stretching & $1178^a$, $1183^d$, $1184^c$\\
& 1369.69 & 1.88$\times$10$^{-18}$  & $CH$ bending &$1393^a$, $1388^d$, $1389^c$\\
& 1409.17 &  3.32$\times$10$^{-18}$  & $NH$ bending &$1409^a$, $1405^d$, $1406^c$\\
& 1417.06 &  1.67$\times$10$^{-17}$  &skeletal deformation &\\
& 1422.89 & 7.68$\times$10$^{-19}$  &$CH_3$ bending&$1463^a$, $1433^d$, $1433^c$ \\
& 1469.28 & 1.33$\times$10$^{-18}$   &  $CH_3$ torsion&$1463^a$, $1455^d$, $1467^c$ \\
& 1490.98 &  3.26$\times$10$^{-18}$   &$CH_2$ scissoring  &\\
& 1499.39 & 9.93$\times$10$^{-18}$    &$CH_2$, $CH_3$ bending &$1518^a$, $1472^d$, $1472^c$ \\
& 1695.39& 8.12$\times$10$^{-18}$    & C=C stretching &$1668^a$, $1684^d$\\ 
& 1752.45 & 1.08$\times$10$^{-16}$   & C=O stretching &$1725^a$, $1712^d$, $1725^c$\\ 
& 1798.79 &  1.34$\times$10$^{-16}$  &C=O stretching&$1772^a$, $1768^d$, $1769^c$\\ 
& 3033.28 & 3.64$\times$10$^{-18}$  & $CH_3$ symmetric stretching &$2984^a$, $2971^d$  \\
& 3085.94 & 1.66$\times$10$^{-18}$ & $CH_2$ symmetric stretching&\\
& 3108.20 & 2.39$\times$10$^{-18}$  &$CH_3$ antisym. stretching  &\\
& 3192.61 & 8.40$\times$10$^{-19}$  & CH stretching &$3076^a$, $2992^d$  \\
& 3596.29 & 1.10$\times$10$^{-17}$   & NH stretching&$3437^a$, $3434^d$, $3434^c$ \\
& 3639.61 & 1.72$\times$10$^{-17}$ & NH stretching&$3484^a$, $3480^d$, $3480^c$ \\
\hline
& 127.65 & 7.98$\times$10$^{-20}$ & CNC bending& \\
& 150.03 & 3.36$\times$10$^{-21}$ & CNC bending& \\ 
& 376.36 & 5.18$\times$10$^{-18}$ & NH bemnding &\\ 
& 385.65& 7.42$\times$10$^{-18}$  & NCC bending &\\ 
& 511.79 & 4.37$\times$10$^{-18}$& NH bending &\\ 
& 518.60 & 1.32$\times$10$^{-18}$& Ring deformation &\\
& 537.81 & 4.50$\times$10$^{-19}$  & Skeletal deformation &\\
& 560.50 & 1.70$\times$10$^{-18}$  & Ring deformation& \\
& 585.53 & 9.75$\times$10$^{-18}$ & NH bending &\\
& 690.21 &  5.27$\times$10$^{-18}$& CCC bending&\\
& 749.65 &1.19$\times$10$^{-17}$ &  C=O bending &\\
& 771.93 & 6.98$\times$10$^{-19}$ & ring deformation&\\
& 785.09 &1.12$\times$10$^{-17}$& CH bending &\\
& 824.37 &4.38$\times$10$^{-18}$ & NH torsion &\\
& 961.25 & 8.90$\times$10$^{-20}$  & CH bending  &\\
& 968.66 & 3.36$\times$10$^{-18}$  & C-C stretching &\\
{\bf $C_4N_3OH_5$}& 988.55 &4.87$\times$10$^{-19}$ & Ring deformation &\\
& 1072.57 & 6.62$\times$10$^{-18}$ & NC stretching &\\
& 1144.44& 4.16$\times$10$^{-18}$ & NH bending& \\
& 1206.67 & 2.36$\times$10$^{-17}$ & CH, NH bending& \\ 
& 1270.31& 1.47$\times$10$^{-17}$ & NC stretching &\\ 
& 1398.63& 4.87$\times$10$^{-18}$  & CH NH torsion &\\ 
& 1420.61& 4.56$\times$10$^{-18}$& NH bending &\\ 
& 1427.43&1.01$\times$10$^{-17}$& C-C stretching &\\
& 1491.01& 1.49$\times$10$^{-17}$  & N-C stretching &\\
& 1665.11& 1.57$\times$10$^{-18}$  & C=c stretching& \\
& 1714.15 & 8.02$\times$10$^{-17}$ & N=C stretching &\\
& 1794.22 &1.33$\times$10$^{-16}$&C=O stretching&\\
& 3208.87&5.19$\times$10$^{-19}$ &  CH stretching &\\
& 3243.95& 2.60$\times$10$^{-19}$ & CH stretching &\\
& 3473.69&1.50$\times$10$^{-18}$& NH stretching &\\
& 3604.26&6.64$\times$10$^{-18}$& NH stretching &\\
& 3646.48 & 1.81$\times$10$^{-17}$  & NH stretching  &\\
\hline
\multicolumn{5}{|c|}{$^a$ Gas phase experiment \citep{cola97}}\\
\multicolumn{5}{|c|}{$^b$ Data recorded in neon matrixes \citep{ivan95}}\\
\multicolumn{5}{|c|}{$^c$ Data recorded in argon matrixes \citep{les92}}\\
\multicolumn{5}{|c|}{$^d$ Data recorded in neon  matrixes \citep{grai90}}\\
\multicolumn{5}{|c|}{$^e$ Data recorded in neon  matrixes \citep{kwia96}}\\
\multicolumn{5}{|c|}{$^f$ Data recorded in argon  matrixes \citep{kwia96}}\\
\hline\hline
\end{longtable}

\end{document}